\documentclass{article}
\usepackage{amsfonts}

\input tcilatex

\begin{document}

\title{Uniformly and nonuniformly elliptic variational equations with gauge
invariance}
\author{Thomas H. Otway\thanks{%
email: otway@ymail.yu.edu} \\
\\
\textit{Department of Mathematics,} \textit{Yeshiva University,}\\
\textit{\ New York, New York 10033}}
\date{}
\maketitle

\begin{abstract}
A large class of variational equations for geometric objects is studied. The
results imply conformal monotonicity and Liouville theorems for steady,
polytropic, ideal flow, and the regularity of weak solutions to generalized
Yang-Mills and Born-Infeld systems. 2000 MSC: 58E15.
\end{abstract}

\section{Introduction}

The Hodge Theorem asserts the existence of a unique harmonic representative
in each cohomology class of $d$-closed forms on a compact Riemannian
manifold, where $d$ is the (flat) exterior derivative. (See, \textit{e.g.}, [%
\textbf{M}, Ch. 7].) This theorem leads to a richly geometric linear model
for stationary fields.\ \ Nonlinear Hodge theory can be viewed as an attempt
to extend the unified geometric interpretation achieved for linear fields to
a large class of quasilinear models.

In the nonlinear generalization of Hodge theory, the set of $d$-closed forms
in $H^{1,2}$ is replaced by a set of $d$-closed forms having finite energy,
the density of which includes a nonlinear function $\rho $ [\textbf{SS1}].\
If $\rho $ is identically 1, this energy functional reduces to the Dirichlet
energy of the linear theory.

Forms of degree 1 occupy a special place in both the linear and nonlinear
Hodge theories, in that $d$-closed 1-forms can be associated to the field of
a scalar potential. In Sec. 2 we consider 2-forms which are closed under 
\textit{covariant} exterior differentiation $D.$ \ These can in certain
circumstances be interpreted as the curvature 2-form derived from a
connection 1-form. The nonlinear Hodge equations in this curved-bundle case
possess additional nonlinearities and nontrivial gauge invariance which are
absent in the conventional flat-bundle case. Sections of the curved bundle
which are stationary points of the nonlinear Hodge energy bear the same
relation to harmonic curvature in a bundle that stationary sections of the
flat bundle bear to harmonic forms on a manifold. \ This leads to a further
generalization of harmonic curvature.

In applications to particle fields a curvature 2-form represents the field
associated to a vector potential. \ In the special case in which the bundle
structure group is the abelian group $U(1)$, $\rho $ can be chosen in such a
way that the nonlinear Hodge energy is equivalent to a suitable
normalization of the abelian Born-Infeld energy. \ This functional is of
interest in connection with string theory interactions [\textbf{Gb}]. \ It
has been studied, partly from a nonlinear Hodge perspective, in [\textbf{Y}%
]. If in addition $\rho \equiv 1,$ then we obtain the Maxwell equations for
time-independent electromagnetic fields on 4-space. If $G$ is nonabelian and 
$\rho \equiv 1,$ then the variational equations of Sec. 2 reduce to the
Yang-Mills equations, the equations for the classical limit of quantum
fields. Additional examples of this kind are given in Sec. 1 of [\textbf{O3}%
].

The material of Sec. 2 is largely expository and is based on [\textbf{O4}].
The new results are in Sec. 3, in which we consider the general case of
section-valued differential forms of arbitrary order, weakly satisfying
nonuniformly elliptic equations associated to the nonlinear Hodge energy. \
The analytic properties derived for such objects are correspondingly weaker
than those derived in Sec. 2. \ The results of Sec. 3 follow from a
conformal monotonicity inequality which we prove for the nonlinear Hodge
energy. The hypotheses of Sec. 3 are satisfied by certain models of
high-speed subsonic flow.

\ In the estimates that follow we denote by $C$ generic positive constants
which generally depend on dimension and which may change in value from line
to line. Dependence of the constants $C$ on bounded variables other than
dimension, where indicated at all, is noted by subscripts.

\section{Lie-algebra-valued sections}

Denote by $X$ a vector bundle over a manifold $M.$ Suppose that $X$ has
compact structure group $G\subset SO(m),$ and that $M$ is a smooth, finite,
oriented, \textit{n}-dimensional Riemannian manifold. Let $A\in \Gamma
\left( M,ad\,X\otimes T^{\ast }M\right) $ be a connection 1-form on $X$
having curvature 2-form 
\[
F_{A}=dA+\frac{1}{2}\left[ A,\,A\right] =dA+A\wedge A, 
\]
where [\ ,\ ] is the bracket of the Lie algebra $\Im ,$ the fiber of the
adjoint bundle $ad\,X.$ Sections of the automorphism bundle $Aut\,X,$ called 
\textit{gauge transformations,} act tensorially on $F_{A}$ but affinely on $%
A $; see, \textit{e.g.,} [\textbf{MM}].

Consider energy functionals having the form 
\begin{equation}
E\left( F_{A}\right) =\int_{M}\left( \int_{0}^{Q}\rho (s)ds\right) dM,
\end{equation}
where $Q=|F_{A}|^{2}=\left\langle F_{A},F_{A}\right\rangle $ is an inner
product on the fibers of the bundle $ad\,X\otimes \Lambda ^{2}\left(
T\;^{\ast }M\right) .$ The inner product on $ad\,X$ is induced by the
normalized trace inner product on $SO(m)$ and that on $\Lambda ^{2}\left(
T\;^{\ast }M\right) ,$ by the exterior product $\ast \left( F_{A}\wedge \ast
F_{A}\right) $, where$\,\ast :\Lambda ^{p}\rightarrow \Lambda ^{n-p}$ is the
Hodge star operator. \ The function $\rho :\Bbb{R}\cup \left\{ 0\right\}
\rightarrow \Bbb{R}^{+}$ is a bounded $C^{1,\alpha }$ function satisfying

\begin{equation}
0<\rho (Q)+2Q\rho ^{\prime }(Q)<\infty
\end{equation}
whenever $Q$ is less than a critical value $Q_{crit}$. \ As $Q$ tends to $%
Q_{crit}$ we retain the right-hand inequality of (2), but allow the middle
term to tend to zero.

The functional (1) appears in \textbf{[SS1]} for the special case $Q=|\omega
|^{2},$ where $\omega \in \Gamma \left( M,\Lambda ^{p}\left( T^{\ast
}M\right) \right) .$ \ In that ``flat-bundle'' case, stationary points with
respect to an admissible cohomology class of closed $p$-forms satisfy the 
\textit{nonlinear Hodge equations} 
\begin{equation}
\delta \left( \rho (Q)\omega \right) =0,
\end{equation}
\begin{equation}
d\omega =0.
\end{equation}

If we choose $X$ to be a bundle having gauge group $U(1),$ making the
choices $p=2$ and 
\[
\rho (Q)=\left( 1+Q\right) ^{-1/2}, 
\]
then $\omega $ has an interpretation as the electromagnetic field of a
(suitably normalized) Born-Infeld energy. In this case condition (2) fails
as $Q$ tends to infinity. \ Equations of nonlinear Hodge type also figure in
elasticity and thermodynamics, including nonrigid-body rotation and
capillarity. Applications to magnetic materials and minimal surfaces are
given in [\textbf{O2}] and [\textbf{SS2}], respectively.

Details on the construction of the nonabelian variational problem are given
in [\textbf{O4}]. The Euler-Lagrange equations for the functional (1) can be
written in the form 
\begin{equation}
\delta \left( \rho (Q)F_{A}\right) =-\ast \left[ A,\ast \rho (Q)F_{A}\right]
,
\end{equation}
where $\delta :\Lambda ^{p}\rightarrow \Lambda ^{p-1}$ is the adjoint of the
exterior derivative $d$. In addition, we have the Bianchi identity 
\begin{equation}
dF_{A}=-\left[ A,F_{A}\right] .
\end{equation}

If $G$ is abelian, then eqs.\ (5) reduce to the system 
\begin{equation}
\delta \left\{ \rho \left[ Q(F_{A})\right] F_{A}\right\} =\delta \left\{
\rho \left[ Q(dA)\right] dA\right\} =0.
\end{equation}
Equations (6) reduce in the abelian case to the equations 
\[
d^{2}A=0, 
\]
which hold automatically on any domain having trivial deRham cohomology.

\subsection{Uniformly elliptic weak solutions}

In this section we derive a H\"{o}lder estimate for weak solutions of the
variational equations. It is easy to show the existence of weak solutions to
(5), (6) by topological arguments, provided that $\rho $ is chosen so that
the energy functional is Palais-Smale. An example is given in Corollary 1.2
of [\textbf{O1}].

\begin{theorem}
Let the pair $(A,F_{A})$ weakly satisfy eqs. (5), (6) in a bounded, open,
Lipschitz domain $\Omega \subset \Bbb{R}^{n},$ $n>2.$ Suppose that there
exist constants $\kappa _{1}$ and $\kappa _{2}$ such that 
\[
0<\kappa _{1}\leq \rho (Q)+2Q\rho ^{\prime }(Q)\leq \kappa _{2}<\infty 
\]
and that $F_{A}\in L^{s}(\Omega )$ for some $s>n/2.$ Then $A$ is equivalent
via a continuous gauge transformation to a connection $\widetilde{A}$ such
that $F_{\widetilde{A}}$ is H\"{o}lder continuous on compact subdomains of $%
\Omega .$
\end{theorem}

\textbf{Remarks.} It is sufficient for $\Omega $ to be \textit{type-A;} see, 
\textit{e.g.,} [\textbf{Gi}]. Theorem 1 was stated and proven in [\textbf{O4}%
]. Here we expand on the main technical point of the proof, the construction
and use of the H\"{o}lder continuous comparison 2-form $d\varphi .$ In doing
so, we modify somewhat the proof in [\textbf{O4}]. In particular, we show
explicitly that a Campanato estimate for $dA$ in a ball can be derived even
if H\"{o}lder estimates for $d\varphi $ do not hold up to the boundary. This
is an important point because, while the analogous estimates can be
continued up to the boundary in the 1-form case [\textbf{O5}], those
arguments fail for higher-degree forms.

\bigskip

\textit{Proof. }Choose coordinates so that the origin lies in the interior
of $\Omega .$ Denote by $B_{r}$ a small $n$-disc of radius $r,$ lying
entirely in the interior of \ $\Omega $ and centered at the origin of
coordinates. Trivialize $X$ locally in order to understand the notion of
weak solution in the sense of [\textbf{Si}, eq. (1.2b)]. For abelian $G$, a 
\textit{weak solution} of (5), (6) is any curvature 2-form $F_{A}$ for which 
$\rho (Q)F_{A}$ is orthogonal in $L^{2}$ to the space of $d$-closed 2-forms $%
d\zeta \in L^{2}(B_{r})$ such that $\zeta \in \Lambda ^{1}$ has vanishing
tangential component on $\partial B.$ For nonabelian $G$, an obvious
extension to inhomogeneous equations allows us to define a weak solution of
(5), (6) by the equation 
\begin{equation}
\int_{B_{r}}\left\langle d\zeta ,\rho (Q)F_{A}\right\rangle \ast
1=-\int_{B_{r}}\left\langle \zeta ,\ast \left[ A,\ast \rho (Q)F_{A}\right]
\right\rangle \ast 1,
\end{equation}
where $F_{A}$ is a curvature 2-form.

We can show $F_{A}$ to be bounded in a smaller concentric ball of radius $%
r/2.$ We do this by extending the arguments of [\textbf{U1}, Sec. 1] to
equations possessing the nonlinear structure of eqs. (5), (6). Using the $%
L^{p}$ hypothesis for $F,$ we carry out the $L^{\infty }$ estimates in a
Hodge gauge. Details are given in [\textbf{O4}, proof of Theorem 1 and the
Appendix (Lemma 7) to Sec. 4].

As gauge transformations act tensorially on $F,$ the curvature remains
bounded under continuous gauge transformations. Choose an exponential gauge
in a euclidean $n$-disc $B_{R},$ $R<r/2,$ centered at the origin of
coordinates in $\Bbb{R}^{n}.$ In such a gauge $A(0)=0$ and $\forall x\in
B_{R},$%
\[
\left| A(x)\right| \leq \frac{1}{2}\left| x\right| \cdot \sup_{\left|
y\right| \leq \left| x\right| }\left| F_{A}(y)\right| ; 
\]
see [\textbf{U2}, Sec. 2]. At the origin of coordinates in an exponential
gauge, $F_{A}$ satisfies eqs. (7). Because $X$ has been trivialized in $B$
we can compare $F_{A}$ to a finite-energy solution $d\varphi $ of the
equation 
\begin{equation}
\int_{B_{R}}\left\langle d\zeta ,\rho \left( \left| d\varphi \right|
^{2}\right) d\varphi \right\rangle \ast 1=0
\end{equation}
for $d\zeta \in L^{2}(B_{R}),$ where $\zeta _{\tan }=0$ on the $\left(
n-1\right) $-sphere $\left| x\right| =R.$ Prescribe vanishing tangential
data for the 1-form $A-\varphi .$ We have 
\[
\int_{B_{R}}\int_{0}^{\left| d\varphi \right| ^{2}}\rho \left( s\right)
ds\ast 1\geq C\int_{B_{R}}\left| d\varphi \right| ^{2}\ast 1 
\]
(\textit{c.f.} [\textbf{U1}, $(1.3)^{\prime }$]), so $d\varphi $ lies in the
space $L^{2}(B_{R})$ by ellipticity and finite energy. In an exponential
gauge $F_{A}=dA,$ so the boundedness of $F_{A}$ implies that $dA$ is
bounded, and thus is certainly in $L^{2}(B_{R}).$ Because $d\left( A-\varphi
\right) $ is in $L^{2},$ we can choose $\zeta =A-\varphi $ in (9). The
resulting weak Dirichlet problem is solvable by Proposition 4.3 of [\textbf{%
Si}]; see also [\textbf{ISS}]. The 2-form $d\varphi $ is H\"{o}lder
continuous in the interior of $B_{R}$ by Proposition 4.4 of [\textbf{Si}],
which is derived from [\textbf{U1}]. \ Thus the Campanato Theorem (Theorem
III.1.2 of [\textbf{Gi}]) implies that 
\[
\int_{B_{R/2}}\left| d\varphi -(d\varphi )_{\left( R/2\right) ,0}\right|
^{2}\ast 1\leq CR^{n+\alpha } 
\]
for some $\alpha \in (0,2],$ where $(f)_{\tau ,\sigma }$ denotes the mean
value of $f$ in an \textit{n}-disc of radius $\tau $ centered at the point $%
\sigma \in \Bbb{R}^{n}.$ \ Combining (8) and (9), we have 
\[
\int_{B_{R}}\left\langle d(A-\varphi ),\rho \left( \left| F_{A}\right|
^{2}\right) F_{A}-\rho \left( \left| d\varphi \right| ^{2}\right) d\varphi
\right\rangle \ast 1= 
\]
\[
-\int_{B_{R}}\left\langle A-\varphi ,\ast \left[ A,\ast \rho \left( \left|
F_{A}\right| ^{2}\right) F_{A}\right] \right\rangle \ast 1. 
\]
The application of a generalized mean-value formula to this equation, as in
Lemma 1.1 of [\textbf{Si}], leads to the inequality 
\[
\int_{B_{R}}\left| d(A-\varphi )\right| ^{2}\ast 1\leq C_{\rho
}(\int_{B_{R}}(|F_{A}|+|d\varphi |)|x|\ast 1+ 
\]
\[
\int_{B_{R}}\left| A-\varphi \right| \left| A\right| \left| F_{A}\right|
\ast 1+\int_{B_{R}}\left| d(A-\varphi )\right| \left| A\right| ^{2}\ast 1) 
\]
\begin{equation}
\equiv C\left( i_{1}+i_{2}+i_{3}\right) .
\end{equation}
Denote by $\varepsilon $ a small, positive number. 
\[
i_{1}=\int_{B_{R}}(|F_{A}|+|d\varphi |)|x|\ast 1\leq
\int_{B_{R}}(|F_{A}|+|d\left( \varphi -A\right) |+\left| dA\right| )|x|\ast
1 
\]
\[
\leq \int_{B_{R}}(|F_{A}|+|dA|)|x|\ast 1+\int_{B_{R}}|d\left( \varphi
-A\right) ||x|\ast 1\leq 
\]
\[
C_{\left\| F_{A}\right\| _{\infty }}\int_{0}^{R}|x|^{n}d\left| x\right|
+\varepsilon \int_{B_{R}}|d\left( \varphi -A\right) |^{2}\ast
1+C_{\varepsilon ,\left| S^{n}\right| }\int_{0}^{R}|x|^{n+1}d\left| x\right|
. 
\]
Denote by $\left\| \cdot \right\| _{p}$ the $L^{p}$-norm over $B_{R}.$ Using
the properties of an exponential gauge, we have by the Sobolev Theorem 
\[
i_{2}=\int_{B_{R}}\left| A-\varphi \right| \left| A\right| \left|
F_{A}\right| \ast 1\leq 
\]
\[
\varepsilon \int_{B_{R}}\left| A-\varphi \right| ^{2}\left| F_{A}\right|
\ast 1+\varepsilon ^{-1}C_{\left\| F_{A}\right\| _{\infty
}}\int_{B_{R}}\left| A\right| ^{2}\ast 1 
\]
\[
\leq \varepsilon \left\| F_{A}\right\| _{n/2}\left\| A-\varphi \right\|
_{2n/\left( n-2\right) }^{2}+C_{\left\| F_{A}\right\| _{\infty },\left|
S^{n}\right| }\int_{0}^{R}|x|^{n+1}d\left| x\right| 
\]
\[
\varepsilon C_{Sobolev,\left\| F_{A}\right\| _{n/2}}\int_{B_{R}}|d\left(
\varphi -A\right) |^{2}\ast 1+C_{\left\| F_{A}\right\| _{\infty }}R^{n+2}, 
\]
and 
\[
i_{3}=\int_{B_{R}}\left| d(A-\varphi )\right| \left| A\right| ^{2}\ast 1\leq 
\]
\[
\varepsilon \int_{B_{R}}\left| d(A-\varphi )\right| ^{2}\ast 1+\varepsilon
^{-1}\int_{B_{R}}\left| A\right| ^{4}\ast 1 
\]
\[
\leq \varepsilon \int_{B_{R}}\left| d(A-\varphi )\right| ^{2}\ast
1+CR^{n+4}. 
\]
Substituting the estimates for $i_{1},$ $i_{2},$ and $i_{3}$ into (10) and
collecting small terms on the left, we obtain 
\[
\int_{B_{R}}\left| d\left( A-\varphi \right) \right| ^{2}\ast 1\leq
CR^{n+1}. 
\]
Then of course 
\[
\int_{B_{R/2}}\left| d\left( A-\varphi \right) \right| ^{2}\ast 1\leq
CR^{n+1}. 
\]
The minimizing property of the mean value with respect to location
parameters implies that 
\[
\int_{B_{R/2}}\left| dA-(dA)_{R,x_{0}}\right| ^{2}\ast 1\leq
\int_{B_{R/2}}\left| dA-(d\varphi )_{R,x_{0}}\right| ^{2}\ast 1 
\]
\[
\leq \int_{B_{R/2}}\left| dA-d\varphi \right| ^{2}\ast
1+\int_{B_{R/2}}\left| d\varphi -(d\varphi )_{R,x_{0}}\right| ^{2}\ast 1 
\]
\begin{equation}
\leq CR^{n+\ell }
\end{equation}
for some $\ell >0.$ Again using the properties of the exponential gauge and
the boundedness of $F_{A},$ we find that we can replace $dA$ by $F_{A}$ in
estimate (11). Because the arguments leading to (11) hold for any radius $%
\tau \in (0,R/2],$ we apply Campanato's Theorem in the form [\textbf{Gi},
Theorem III.1.3] to conclude that $F_{A}$ is H\"{o}lder continuous in $%
B_{R/2}.$

We have used the exponential gauge at the origin of coordinates. In order to
remove this limitation, suppose that a map $\gamma \in AutX$ is continuous
at each point $x\in B_{\tau }(\sigma ),$ an \textit{n}-disc of sufficiently
small radius $\tau $ centered at a point $\sigma .$ Using the continuity of $%
\gamma \ $and the fact that $\gamma $ is unitary, we have for small $r,$%
\[
\left\| \gamma ^{-1}(x)F_{A}(x)\gamma (x)-\left[ \gamma
^{-1}(x)F_{A}(x)\gamma (x)\right] _{\tau ,\sigma }\right\| _{2}\approx 
\]
\[
\left\| \gamma ^{-1}(x)F_{A}(x)\gamma (x)-\left[ \gamma ^{-1}(\sigma
)F_{A}(x)\gamma (\sigma )\right] _{\tau ,\sigma }\right\| _{2}\leq 
\]
\[
\left\| F_{A}(x)\left( \gamma (x)\gamma ^{-1}(\sigma )-I\right) +\left(
I-\gamma (x)\gamma ^{-1}(\sigma )\right) \left[ F_{A}(x)\right] _{\tau
,\sigma }\right\| _{2} 
\]
\[
+\left\| F_{A}(x)-\left[ F_{A}(x)\right] _{\tau ,\sigma }\right\| _{2}, 
\]
where $I$ is the identity transformation and the $L^{2}$-norms are taken
over $B_{\tau }\left( \sigma \right) .$ This implies the inequality 
\[
\left\| \gamma ^{-1}(x)F_{A}(x)\gamma (x)-\left[ \gamma
^{-1}(x)F_{A}(x)\gamma (x)\right] _{r,\sigma }\right\| _{2}\leq 
\]
\[
\left\| \left( \gamma (x)\gamma ^{-1}(\sigma )-I\right) \left( F_{A}(x)- 
\left[ F_{A}(x)\right] _{r,\sigma }\right) \right\| _{2}+Cr^{\left(
n+1\right) /2}\leq C^{\prime }r^{\left( n+1\right) /2}, 
\]
which allows us to complete the proof of Theorem 1 by a covering argument.

\section{A conformal monotonicity formula}

In this section we extend the results of [\textbf{O2}] concerning conformal
monotonicity properties of 2-forms satisfying an ellipticity condition to
forms of arbitrary degree. In place of the ellipticity condition on the
variational equations, we impose pointwise monotonicity and noncavitation
hypotheses on the energy density. A standard construction [\textbf{P}]
allows the result to be applied to solutions of the gauge-invariant
equations studied in the preceding section.

Recall that a functional is said to be \textit{r-stationary} [\textbf{A}] if
it is stationary with respect to compactly supported $C^{1}$
reparametrizations of its domain.

\begin{theorem}
Let $\omega $ be a form of order $q\geq 1$ and let the associated nonlinear
Hodge energy $E(\omega )$ be r-stationary on a domain $\Omega $ of $\Bbb{R}%
^{n}$ containing the unit $n$-disc, $n>2q$. \ Suppose that the (conformally
weightless) energy density $\rho $ satisfies $\rho ^{\prime }(s)\leq
0\;\forall s\in \lbrack 0,s_{crit}].$ Then for almost every point of $\Omega 
$ and for $0<r_{1}<r_{2},$ we have 
\[
r_{1}^{2q-n}E_{|B_{r_{1}}}\leq r_{2}^{2q-n}E_{|B_{r_{2}}}, 
\]
where $B_{r}$ is an $n$-disc, of radius $r,$ completely contained in the
interior of $\Omega .$
\end{theorem}

\textbf{Remarks}. If $q=1,$ then Theorem 2 has an interpretation as a
monotonicity formula for steady, polytropic, ideal flow. The theorem is true
for \ the case in which $\omega $ is the differential of a map into a
Riemannian manifold $N$; see, \textit{e.g.}, [\textbf{T}] for a proof in the
constant-density case. \ If $\omega $ is the curvature of a
Lie-algebra-valued connection 1-form, the result holds for suitably lifted
r-variations of $\omega ;$ this is illustrated by [\textbf{P}] in the
constant-density case and [\textbf{O2}] in the case of nonlinear mass
density. The flatness of $\Omega $ is of somewhat more than notational
significance: \ this hypothesis simplifies the proof by permitting the use
of the expansion (12); but it is otherwise of little importance \textit{(c.f.%
} [\textbf{P}]). \ Certain other hypotheses of Theorem 2 can be weakened. \
For example, we will show the theorem to be true if $E(\omega )$ is
r-stationary only on $\Omega /\Sigma ,$ where $\Sigma $ is a compact subset
of sufficiently small Hausdorff dimension (Theorem 6). The theorem also
remains true if $E(\omega )$ is not quite r-stationary but satisfies a
certain integral inequality on $\Omega /\Sigma ;$ see Sec. 2 of [\textbf{O2}%
] for the special case of 2-forms. \ Because [\textbf{O2}] places no
assumption on the sign of $\rho ^{\prime }$, that result depends on the
ellipticity coefficients, which our result does not.

\bigskip

\textit{Proof of Theorem 2.} \ Denote by $\psi ^{t}$ a 1-parameter family of
compactly supported diffeomorphisms of $\Omega $ such that 
\[
\psi ^{s}\circ \psi ^{t}=\psi ^{s+t},\; 
\]
and\ $\psi ^{0}=identity.$ \ The r-variations of $E(\omega )$ are given by 
\[
\delta _{r}E(\omega )=\frac{d}{dt}_{|t=0}E\left( \psi ^{t\ast }\omega
\right) . 
\]
The first step of the proof is to obtain an explicit expression for this
quantity.

Write 
\begin{equation}
f\equiv \psi ^{t}(x)=x+t\xi (x)+O(t^{2}),
\end{equation}
where 
\[
\xi (x)=\frac{d}{dt}_{|t=0}\psi ^{t}\left( x\right) 
\]
is \textit{the variation vector field}, which will be chosen later in the
proof.

\begin{lemma}
\[
\frac{d}{dt}_{|t=0}\omega _{i_{1}\cdots i_{q}}(f)df^{i_{1}}\cdots
df^{i_{q}}= 
\]
\[
\frac{d}{dt}_{|t=0}\omega _{i_{1}\cdots i_{q}}(f)A_{\ell _{1}\cdots \ell
_{q}}^{i_{1}\cdots i_{q}}(q,\xi ,x,t)dx^{\ell _{1}}\cdots dx^{\ell _{q}}, 
\]
where 
\[
A_{\ell _{1}\cdots \ell _{q}}^{i_{1}\cdots i_{q}}(q,\xi ,x,t)=\delta _{\ell
_{1}}^{i_{1}}\cdots \delta _{\ell _{q}}^{i_{q}}+\sum_{j=1}^{q}\left( \delta
_{\ell _{1}}^{i_{1}}\cdots \widehat{\delta _{\ell _{j}}^{i_{j}}}\cdots
\delta _{\ell _{q}}^{i_{q}}\right) t\frac{\partial \xi ^{i_{j}}}{\partial
x^{\ell _{j}}}+O\left( t^{2}\right) . 
\]
\end{lemma}

We prove Lemma 3 by induction on the case $q=2$, the simplest nontrivial
case. We have 
\[
\frac{d}{dt}_{|t=0}\omega _{ij}(f)df^{i}df^{j}=\frac{d}{dt}_{|t=0}\omega
_{ij}(f)\frac{\partial f^{i}}{\partial x^{k}}dx^{k}\frac{\partial f^{j}}{%
\partial x^{m}}dx^{m}= 
\]
\[
\frac{d}{dt}_{|t=0}\omega _{ij}(f)\left( \delta _{k}^{i}+t\frac{\partial \xi
^{i}}{\partial x^{k}}\right) dx^{k}\left( \delta _{m}^{j}+t\frac{\partial
\xi ^{j}}{\partial x^{m}}\right) dx^{m}= 
\]
\begin{equation}
\frac{d}{dt}_{|t=0}\omega _{ij}(f)\left( \delta _{k}^{i}\delta
_{m}^{j}+\delta _{k}^{i}t\frac{\partial \xi ^{j}}{\partial x^{m}}+\delta
_{m}^{j}t\frac{\partial \xi ^{i}}{\partial x^{k}}+O(t^{2})\right)
dx^{k}dx^{m}.
\end{equation}
Assume that eq.\ (13) holds for $q=m$. \ We show that the formula must also
hold for $q=m+1$. \ We have 
\[
\frac{d}{dt}_{|t=0}\omega _{i_{1}\cdots i_{m+1}}(f)df^{i_{1}}\cdots
df^{i_{m+1}}= 
\]
\[
\frac{d}{dt}_{|t=0}\omega _{i_{1}\cdots i_{m+1}}(f)\left( \delta _{\ell
_{1}}^{i_{1}}+t\frac{\partial \xi ^{i_{1}}}{\partial x^{\ell _{1}}}\right)
dx^{\ell _{1}}\cdots \left( \delta _{\ell _{m+1}}^{i_{m+1}}+t\frac{\partial
\xi ^{i_{m+1}}}{\partial x^{\ell _{m+1}}}\right) dx^{\ell _{m+1}}= 
\]
\[
\frac{d}{dt}_{|t=0}\omega _{i_{1}\cdots i_{m+1}}(f)A_{\ell _{1}\cdots \ell
_{m+1}}^{i_{1}\cdots i_{m+1}}(m,\xi ,x,t)dx^{\ell _{1}}\cdots dx^{\ell
_{m+1}}. 
\]
By the induction hypothesis 
\[
A_{\ell _{1}\cdots \ell _{m}}^{i_{1}\cdots i_{m}}(m,\xi ,x,t)= 
\]
\[
\lbrack \delta _{\ell _{1}}^{i_{1}}\cdots \delta _{\ell _{m}}^{i_{m}}+t\frac{%
\partial \xi ^{i_{1}}}{\partial x^{\ell _{1}}}\delta _{\ell
_{2}}^{i_{2}}\cdots \delta _{\ell _{m}}^{i_{m}}+\sum_{j=2}^{m-1}\left(
\delta _{\ell _{1}}^{i_{1}}\cdots \widehat{\delta _{\ell _{j}}^{i_{j}}}%
\cdots \delta _{\ell _{m}}^{i_{m}}\right) t\frac{\partial \xi ^{i_{j}}}{%
\partial x^{\ell _{j}}} 
\]
\[
+\delta _{\ell _{1}}^{i_{1}}\cdots \delta _{\ell _{m-1}}^{i_{m-1}}t\frac{%
\partial \xi ^{i_{m}}}{\partial x^{\ell _{m}}}+O\left( t^{2}\right) ]\left(
\delta _{\ell _{m+1}}^{i_{m+1}}+t\frac{\partial \xi ^{i_{m+1}}}{\partial
x^{\ell _{m+1}}}\right) = 
\]
\[
\delta _{\ell _{1}}^{i_{1}}\cdots \delta _{\ell _{m+1}}^{i_{m+1}}+[t\frac{%
\partial \xi ^{i_{1}}}{\partial x^{\ell _{1}}}\delta _{\ell
_{2}}^{i_{2}}\cdots \delta _{\ell _{m}}^{i_{m}}+\sum_{j=2}^{m-1}\left(
\delta _{\ell _{1}}^{i_{1}}\cdots \widehat{\delta _{\ell _{j}}^{i_{j}}}%
\cdots \delta _{\ell _{m}}^{i_{m}}\right) t\frac{\partial \xi ^{i_{j}}}{%
\partial x^{\ell _{j}}}+ 
\]
\[
\delta _{\ell _{1}}^{i_{1}}\cdots \delta _{\ell _{m-1}}^{i_{m-1}}t\frac{%
\partial \xi ^{i_{m}}}{\partial x^{\ell _{m}}}]\delta _{\ell
_{m+1}}^{i_{m+1}}+\delta _{\ell _{1}}^{i_{1}}\cdots \delta _{\ell
_{m}}^{i_{m}}t\frac{\partial \xi ^{i_{m+1}}}{\partial x^{\ell _{m+1}}}%
+O\left( t^{2}\right) = 
\]
\[
\delta _{\ell _{1}}^{i_{1}}\cdots \delta _{\ell _{m+1}}^{i_{m+1}}+t\frac{%
\partial \xi ^{i_{1}}}{\partial x^{\ell _{1}}}\delta _{\ell
_{2}}^{i_{2}}\cdots \delta _{\ell _{m+1}}^{i_{m+1}}+\sum_{j=2}^{m}\left(
\delta _{\ell _{1}}^{i_{1}}\cdots \widehat{\delta _{\ell _{j}}^{i_{j}}}%
\cdots \delta _{\ell _{m+1}}^{i_{m+1}}\right) t\frac{\partial \xi ^{i_{j}}}{%
\partial x^{\ell _{j}}} 
\]
\[
+\delta _{\ell _{1}}^{i_{1}}\cdots \delta _{\ell _{m}}^{i_{m}}t\frac{%
\partial \xi ^{i_{m+1}}}{\partial x^{\ell _{m+1}}}+O\left( t^{2}\right) . 
\]
This proves the lemma.

\bigskip

Make the coordinate transformation $x\rightarrow y,$ where 
\[
y=\left( \psi ^{t}\right) ^{-1}(x). 
\]

\begin{lemma}
In terms of $y$, 
\[
\frac{d}{dt}_{|t=0}\omega _{i_{1}\cdots i_{q}}(f)df^{i_{1}}\cdots
df^{i_{q}}= 
\]
\[
\omega _{i_{1}\cdots i_{q}}(x)\sum_{j=1}^{q}\frac{\partial \xi ^{i_{j}}}{%
\partial x^{\ell _{j}}}dx^{\ell _{1}}\cdots \widehat{dx^{\ell _{j}}}\cdots
dx^{\ell _{q}}= 
\]
\begin{equation}
q\omega _{i_{1}i_{2}\cdots i_{q}}(x)\frac{\partial \xi ^{i_{1}}}{\partial
x^{\ell _{1}}}dx^{\ell _{1}}dx^{\ell _{2}}\cdots dx^{\ell _{q}}.
\end{equation}
\end{lemma}

\textit{Proof.} \ Notice that if $q=2$, 
\[
\frac{d}{dt}_{|t=0}\omega _{ij}(f)df^{i}df^{j}=\omega _{ij}(x)\left( \frac{%
\partial \xi ^{j}}{\partial x^{m}}dx^{i}dx^{m}+\frac{\partial \xi ^{i}}{%
\partial x^{k}}dx^{k}dx^{j}\right) = 
\]
\begin{equation}
2\omega _{ij}(x)\frac{\partial \xi ^{i}}{\partial x^{k}}dx^{k}dx^{j},
\end{equation}
where in the last identity we used the fact that both $\omega _{ij}$ and $%
dx^{i}dx^{j}$ are antisymmetric in $i$ and $j$. \ We show that this
reasoning extends, with mainly notational complications, to forms of
arbitrary order.

Lemma 3 implies that 
\[
\frac{d}{dt}_{|t=0}\omega _{i_{1}\cdots i_{q}}(f)df^{i_{1}}\cdots
df^{i_{q}}= 
\]
\[
\omega _{i_{1}\cdots i_{q}}[\frac{\partial \xi ^{i_{1}}}{\partial x^{m}}%
dx^{m}dx^{i_{2}}\cdots dx^{i_{q}}+\sum_{j=2}^{q-1}dx^{i_{1}}\cdots \frac{%
\partial \xi ^{i_{j}}}{\partial x^{m}}dx^{m}\cdots dx^{i_{q}} 
\]
\begin{equation}
+dx^{i_{1}}\cdots dx^{i_{q-1}}\frac{\partial \xi ^{i_{q}}}{\partial x^{m}}%
dx^{m}]\equiv i_{1}+i_{2}+i_{3}.
\end{equation}
Here 
\[
i_{1}+i_{3}=\omega _{i_{1}\cdots i_{q}}\frac{\partial \xi ^{i_{1}}}{\partial
x^{m}}dx^{m}\wedge \left( dx^{i_{2}}\cdots dx^{i_{q}}\right) + 
\]
\[
\omega _{i_{1}\cdots i_{q}}\left( dx^{i_{1}}\cdots dx^{i_{q-1}}\right)
\wedge \frac{\partial \xi ^{i_{q}}}{\partial x^{m}}dx^{m}=i_{1}+ 
\]
\begin{equation}
\omega _{i_{q}i_{1}\cdots i_{q-1}}\frac{\partial \xi ^{i_{q}}}{\partial x^{m}%
}dx^{m}\wedge \left( dx^{i_{1}}\cdots dx^{i_{q-1}}\right) =2i_{1}
\end{equation}
by a relabelling of indices in the second term of the sum. 
\[
i_{2}=\omega _{i_{1}\cdots i_{q}}\sum_{j=2}^{q-1}\left( dx^{i_{1}}\cdots
dx^{i_{j-1}}\right) \wedge \frac{\partial \xi ^{i_{j}}}{\partial x^{m}}%
dx^{m}\wedge dx^{i_{j+1}}\cdots dx^{i_{q}}= 
\]
\[
-\omega _{i_{1}\cdots i_{j-1}i_{j}i_{j+1}\cdots i_{q}}\sum_{j=2}^{q-1}\frac{%
\partial \xi ^{i_{j}}}{\partial x^{m}}dx^{m}\wedge \left( dx^{i_{1}}\cdots
dx^{i_{j-1}}\right) \wedge \left( dx^{i_{j+1}}\cdots dx^{i_{q}}\right) = 
\]
\[
\omega _{i_{j}i_{1}\cdots \widehat{i_{j}}\cdots i_{q}}\sum_{j=2}^{q-1}\frac{%
\partial \xi ^{i_{j}}}{\partial x^{m}}dx^{m}\wedge \left( dx^{i_{1}}\cdots
dx^{i_{j-1}}\right) \wedge \left( dx^{i_{j+1}}\cdots dx^{i_{q}}\right) 
\]
\begin{equation}
=(q-2)\omega _{i_{1}\cdots i_{q}}\frac{\partial \xi ^{i_{1}}}{\partial x^{m}}%
dx^{m}dx^{i_{2}}\cdots dx^{i_{q}}=\left( q-2\right) i_{1}.
\end{equation}
The last identity results from a relabelling of indices. \ Substituting
identities (17) and (18) into the right-hand side of (16) extends (15) to
the case of arbitrary $q$. \ This proves Lemma 4.

\bigskip

If $J$ is the Jacobian of the transformation $x\rightarrow y,$ then from
(12) we obtain 
\begin{equation}
\frac{d}{dt}_{|t=0}J\left[ \left( \psi ^{t}\right) ^{-1}\right] =\frac{d}{dt}%
_{|t=0}\left| \frac{\partial x}{\partial f}\right| =-div\,\xi .
\end{equation}
Define 
\[
e(Q)=\int_{0}^{Q}\rho (s)\,ds. 
\]
By hypothesis, 
\[
0=\delta _{r}E(\omega )=\frac{d}{dt}_{|t=0}\int_{\Omega }e\left(
\left\langle \psi ^{t\ast }\omega ,\psi ^{t\ast }\omega \right\rangle
\right) \,J\left[ \left( \psi ^{t}\right) ^{-1}\right] \,\ast 1= 
\]
\[
\int_{\Omega }e(Q)\frac{d}{dt}_{|t=0}J\left[ \left( \psi ^{t}\right) ^{-1}%
\right] \ast 1+ 
\]

\ 
\begin{equation}
\int_{\Omega }e^{\prime }(Q)2\left\langle \frac{d}{dt}_{|t=0}\omega
_{i_{1}\cdots i_{q}}(f)df^{i_{1}}\cdots df^{i_{q}},\omega _{\ell _{1}\cdots
\ell _{q}}(f)df^{\ell _{1}}\cdots df^{\ell _{q}}\right\rangle \ast 1.
\end{equation}
(The weightlessness of $\rho $ is used on the right-hand side of this
expression.) Substituting (14) and (19) into (20) yields 
\[
\int_{\Omega }e(Q)\,div\,\xi \ast 1= 
\]
\begin{equation}
2q\int_{\Omega }e^{\prime }(Q)\left\langle \omega _{i_{1}i_{2}\cdots
i_{q}}(x)\frac{\partial \xi ^{i_{1}}}{\partial x^{\ell _{j}}}dx^{\ell
_{j}}dx^{\ell _{2}}\cdots dx^{\ell _{q}},\omega _{i_{1}\cdots
i_{q}}(f)df^{i_{1}}\cdots df^{i_{q}}\right\rangle \ast 1.
\end{equation}

Choose an orthonormal basis 
\[
\left\{ u_{_{i}}\right\} _{i=1}^{n}=\left\{ \frac{\partial }{\partial r},%
\frac{\partial }{\partial \theta _{2}},\ldots ,\frac{\partial }{\partial
\theta _{n}}\right\} 
\]
and let [\textbf{P}] 
\[
\xi =\eta (r)r\cdot \frac{\partial }{\partial r}, 
\]
where $r$ is the radial coordinate in a curvilinear system; $\eta (r)\in
C_{0}^{\infty }\left[ 0,1\right] ;$ $\eta ^{\prime }(r)\leq 0;$\ $\eta
(r)=v\left( r/\tau \right) =1$ for $r\leq \tau ,$ where $\tau $ is a number
in the interval $(0,1);$ there is a positive number $\delta $ for which $%
\eta (r)=0$ whenever $r$ exceeds $\tau +\delta .$ \ For this choice of $\xi $%
, 
\[
div\,\xi =\nabla _{\frac{\partial }{\partial r}}\cdot \xi +\nabla _{\frac{%
\partial }{\partial \theta _{m}}}\cdot \xi , 
\]
$m=2,\ldots ,n.$ But 
\[
\nabla _{\frac{\partial }{\partial \theta _{m}}}\cdot \xi =\left[ r\eta (r)%
\right] \nabla _{\frac{\partial }{\partial \theta _{m}}}\cdot \frac{\partial 
}{\partial r}= 
\]
\[
\eta (r)\nabla _{r\frac{\partial }{\partial \theta _{m}}}\cdot \frac{%
\partial }{\partial r}=\sum_{m=2}^{n}\eta (r)\frac{\partial }{\partial
\theta _{m}}, 
\]
so 
\[
div\,\xi =\left( r\eta \right) ^{\prime }+\left( n-1\right) \eta . 
\]

Initially, let $q=2.$ We compute, for our choice of $\xi ,$ and $%
i,j=1,...,n, $%
\[
\left\langle \omega \left( \nabla _{i}\xi ,u_{j}\right) ,\omega
(u_{i},u_{j})\right\rangle =\frac{d}{dr}\left( r\eta \right) \left\langle
\omega \left( \frac{\partial }{\partial r},\frac{\partial }{\partial \theta
_{m}}\right) ,\omega \left( \frac{\partial }{\partial r},\frac{\partial }{%
\partial \theta _{m}}\right) \right\rangle 
\]
\begin{equation}
+\eta (r)\left\langle \omega \left( \frac{\partial }{\partial \theta _{\ell }%
},\frac{\partial }{\partial \theta _{m}}\right) ,\omega \left( \frac{%
\partial }{\partial \theta _{\ell }},\frac{\partial }{\partial \theta _{m}}%
\right) \right\rangle _{|\ell \neq m},
\end{equation}
where $\ell ,m=2,\ldots ,n;$ we have used in computing eq.\ (22) the facts
that $dr\wedge dr=0$ and the basis vectors are orthonormal. \ Evaluating $%
\left( r\eta \right) ^{\prime }$ by the product rule, we can write eq.\ (21)
in the form 
\[
\int_{\Omega }e(Q)\left[ \eta +r\eta ^{\prime }+(n-1)\eta \right] \,\ast 1= 
\]
\[
4\int_{\Omega }\rho (Q)\left\{ \left( \eta +r\eta ^{\prime }\right)
\sum_{m=2}^{n}\left| \omega \left( \frac{\partial }{\partial r},\frac{%
\partial }{\partial \theta _{m}}\right) \right| ^{2}+\eta \sum_{\ell
,m=2,\ell \neq m}^{n}\left| \omega \left( \frac{\partial }{\partial \theta
_{\ell }},\frac{\partial }{\partial \theta _{m}}\right) \right| ^{2}\right\}
\,\ast 1. 
\]
The extension to arbitrary $q$ follows by obvious notational alterations for 
$\omega =\omega \left( u_{i_{1}},\ldots ,u_{i_{q}}\right) .$

We thus obtain 
\[
\int_{\Omega }e(Q)\,\left( n\eta +r\eta ^{\prime }\right) \ast 1= 
\]
\begin{equation}
2q\int_{\Omega }Q\rho (Q)\eta \,\ast 1+2q\int_{\Omega }\rho (Q)r\eta
^{\prime }\left| \frac{\partial }{\partial r}\rfloor \omega \right|
^{2}\,\ast 1,
\end{equation}
Our hypothesis on the sign of $\rho ^{\prime }$ implies that 
\[
Q\rho (Q)=\int_{0}^{Q}\frac{d}{ds}\left( s\rho (s)\right) \,ds 
\]
\begin{equation}
=\int_{0}^{Q}\left[ s\rho (s)\rho ^{\prime }(s)+\rho (s)\right] \,ds\leq
\int_{0}^{Q}\rho (s)\,ds=e\left( Q\right) .
\end{equation}
Substituting this into (23) yields 
\[
\int_{\Omega }e(Q)\,\left( n\eta -2q\eta +r\eta ^{\prime }\right) \ast 1\leq 
\]
\begin{equation}
2q\int_{\Omega }\rho (Q)r\eta ^{\prime }\left| \frac{\partial }{\partial r}%
\rfloor \omega \right| ^{2}\,\ast 1.
\end{equation}
By construction 
\[
r\eta ^{\prime }(r)=-\tau \frac{\partial }{\partial \tau }v\left( \frac{r}{%
\tau }\right) \leq 0. 
\]
This yields 
\[
0\leq 2q\int_{\Omega }\rho (Q)\tau \frac{\partial }{\partial \tau }v\left( 
\frac{r}{\tau }\right) \left| \frac{\partial }{\partial r}\rfloor \omega
\right| ^{2}\,\ast 1\leq 
\]
\[
\int_{\Omega }e(Q)\left[ \left( 2q-n\right) \eta +\tau \frac{\partial }{%
\partial \tau }v\left( \frac{r}{\tau }\right) \right] \,\ast 1. 
\]
As $\delta $ tends to zero we obtain 
\[
0\leq \left( 2q-n+\tau \frac{\partial }{\partial \tau }\right) \int_{B_{\tau
}}e(Q)\,\ast 1. 
\]
Multiplying this last inequality by the integrating factor $\tau
^{2q-(n+1)}, $ the proof of Theorem 2 is completed by integration over $\tau 
$ between $r_{1}$ and $r_{2}$.

\subsection{An application and an extension}

\begin{corollary}
Assume the hypotheses of Theorem 2 for $\Omega =\Bbb{R}^{n}$ and suppose
that 
\[
E_{|B_{r}}\leq Cr^{k} 
\]
as $r$ tends to infinity for sufficiently small $k.$\ Suppose that $\rho (Q)$
is bounded below away from zero. \ Then $Q(x)$ is zero for almost every $%
x\in \Omega .$
\end{corollary}

This result was proven for the case $\rho \equiv 1$ in [\textbf{P}].

\bigskip

\textit{Proof.} \ Without loss of generality, take $k$ to be nonnegative. We
can write the growth condition in the form 
\begin{equation}
r^{2q-n}E_{|B_{r}}\leq Cr^{2q+k-n},
\end{equation}
where $2q+k-n<0$ for sufficiently small $k$. \ The right-hand side of (26)
tends to zero as $r$ tends to infinity. \ The left-hand side is nonnegative
by construction. \ Thus the conformal energy $r^{2q-n}E_{|B_{r}}$ tends to
zero on $\Bbb{R}^{n}$. \ Because by Theorem 2 the conformal energy is
nondecreasing for increasing $r$, we conclude that $E$ is identically zero
on $\Bbb{R}^{n}$. \ The vanishing of the energy on a ball of infinite radius
implies the pointwise vanishing of $Q$ almost everywhere by the inequality 
\[
\frac{1}{2}\int_{\Omega }\int_{0}^{Q}\rho (s)\,ds\,\ast 1\geq \frac{1}{2}%
\min_{s\in \left[ 0,Q\right] }\rho (s)\int_{\Omega }\int_{0}^{Q}ds\,\ast
1\geq C\int_{\Omega }Q\,\ast 1, 
\]
which follows from our assumption that $\rho (Q)$ is bounded below away from
0. \ For example, if $\omega $ is the 1-form canonically associated by the
inner product to the velocity of a steady, polytropic, ideal flow we have,
by noncavitation, 
\[
\int_{0}^{Q}\rho (s)\,ds=\frac{2}{\gamma }\left[ 1-\left( 1-\frac{\gamma -1}{%
2}Q\right) ^{\gamma /\left( \gamma -1\right) }\right] 
\]
\[
\geq \frac{2}{\gamma }\left[ 1-\left( 1-\frac{\gamma -1}{2}Q\right) \right] =%
\frac{\gamma -1}{\gamma }Q, 
\]
where $\gamma >1$ is the adiabatic constant of the medium. \ This inequality
holds even for transonic flow, provided $Q$ is exceeded by the number $%
2/\left( \gamma -1\right) .$

\begin{theorem}
Let $\Sigma $ be a compact singular set, of codimension $m\in (2,n],$
completely contained in a sufficiently small ball which is itself completely
contained in the interior of $\Omega ,$ such that $\partial \Sigma $ is
Lipschitz. \ If the domain $\Omega $ is replaced by the domain $\Omega
/\Sigma $ in Theorem 2 and $\Bbb{R}^{n}$ is replaced by the domain $\Bbb{R}%
^{n}/\Sigma $ in Corollary 5, then the assertions of these propositions
continue to hold provided we add the hypothesis that $Q(\omega )\in
L^{m/\left( m-1\right) }(\Omega ).$\ \ If $\Sigma $ is a point, then the $%
L^{p}$ condition on $Q$ can be replaced by a hypothesis of finite energy.
\end{theorem}

\textbf{Remark}. \ For certain choices of $\rho $ the $L^{p}$ condition on $%
Q $ can of course be obtained from finite energy. \ If $\partial \Sigma $ is
not Lipschitz, then the result is true under a slightly stronger $L^{p}$
hypothesis (\textit{c.f.} [\textbf{O2}]).

\textit{Proof}. \ Replace the variation vector field in the proof of Theorem
1 by the quantity [\textbf{O2}] 
\begin{equation}
\xi =\left( 1-\chi ^{(\nu )}\right) \eta (r)r\cdot \left( \partial /\partial
r\right) ,
\end{equation}
where $\chi ^{(\nu )}$ denotes a sequence of functions of $r$ such that $%
\chi ^{(\nu )}\in \lbrack 0,1]$ and $\chi ^{(\nu )}$ is equal to 1 in a
neighborhood of the singular set $\Sigma .$ \ If $\Sigma $ has zero $s$%
-capacity with respect to $\Omega $ for $1\leq s\leq n,$ then $\chi ^{(\nu
)} $ can be chosen so that as $\nu $ tends to infinity, $\chi ^{(\nu
)}\rightarrow 0$ a.e. and $\nabla \chi ^{(\nu )}\rightarrow 0$ in $L^{s}$ [%
\textbf{Se}, Lemma 2 and p. 73]. \ Because $\rho $ is noncavitating, we have
by (24), 
\begin{equation}
\int_{\Omega }e(Q)\chi ^{(\nu )\prime }(r)r\eta \ast 1\geq -C\left\| \nabla
\chi ^{(\nu )}\right\| _{L^{m}}\left\| Q\right\| _{L^{m/(m-1)}}.
\end{equation}
By a result of Carlson [\textbf{W}], a Lipschitz set of codimension $m$ has
vanishing $m$-capacity, so the right-hand side of this inequality tends to
zero as $\nu $ tends to infinity. \ Similarly, 
\begin{equation}
\int_{\Omega }\rho (Q)\chi ^{(\nu )\prime }(r)r\eta \left| \frac{\partial }{%
\partial r}\rfloor \omega \right| ^{2}\ast 1\leq C\left\| \nabla \chi ^{(\nu
)}\right\| _{L^{m}}\left\| Q\right\| _{L^{m/(m-1)}}.
\end{equation}
The analogue of the left-hand and right-hand sides of inequality (25) that
arises from taking the variation vector field given to equal (27) will
contain terms that can be estimated by inequalities (28) and (29),
respectively. \ This completes the proof of Theorem 6 in the case $\dim
(\Sigma )>0$.

If $\Sigma $ is a point, then we can choose coordinates in which the
singularity lies at the origin of $\Bbb{R}^{n}$ and define [\textbf{L}] 
\begin{equation}
\xi =\zeta \eta (r)r\cdot \left( \partial /\partial r\right) ,
\end{equation}
where $\zeta =0$ in a ball $B_{\sigma }$ of radius $\sigma $ about the
singularity, $\zeta \left( \left| x\right| \right) =1$ for $\left| x\right| $
exceeding $2\sigma ,$ and 
\begin{equation}
\zeta ^{\prime }(\left| x\right| )\leq C\sigma ^{-1}
\end{equation}
for $x\in B_{2\sigma }/B_{\sigma }$. \ We obtain 
\begin{equation}
\int_{\Omega }e(Q)\zeta ^{\prime }(r)r\eta \ast 1\geq -C\int_{B_{2\sigma
}/B_{\sigma }}e(Q)\eta \ast 1
\end{equation}
and 
\[
\int_{\Omega }\rho (Q)\zeta ^{\prime }(r)r\eta \left| \frac{\partial }{%
\partial r}\rfloor \omega \right| ^{2}\ast 1\leq C\int_{B_{2\sigma
}/B_{\sigma }}\rho (Q)\eta \left| \frac{\partial }{\partial r}\rfloor \omega
\right| ^{2}\ast 1 
\]
\begin{equation}
\leq C\int_{B_{2\sigma }/B_{\sigma }}\rho (Q)\eta Q\ast 1\leq
C\int_{B_{2\sigma }/B_{\sigma }}e(Q)\eta \ast 1.
\end{equation}
Locally finite energy implies that the right-hand sides of both (32) and
(33) tend to zero as $\sigma $ tends to zero. \ Considering these
inequalities in evaluating the extra terms introduced into (25) by choosing $%
\xi $ as in (30) and (31) completes the proof for the case $\dim (\Sigma )=0$%
.

This completes the proof of Theorem 6.

\bigskip

\[
\mathbf{References} 
\]

[\textbf{A}] \ W. Allard, On the first variation of a varifold, \textit{Ann.
Math.} \textbf{95} (1972), 417-491.

[\textbf{Gi]} \ M. Giaquinta, \textit{Multiple Integrals in the Calculus of
Variations and Nonlinear Elliptic Systems,} Princeton University Press,
Princeton, 1983.

[\textbf{Gb}] \ G. W. Gibbons, Born-Infeld particles and Dirichlet $p$%
-branes, \textit{Nucl. Phys}. \textbf{B 514 }(1998), 603-639.

[\textbf{ISS}] T. Iwaniec, C. Scott, and B. Stroffolini, Nonlinear Hodge
theory on manifolds with boundary, \textit{Annali Mat. Pura Appl.} (4) 
\textbf{177} (1999), 37-115.

[\textbf{L}] \ G. Liao, A regularity theorem for harmonic maps with small
energy, \textit{J. Differential Geometry} \textbf{22} (1985), 233-241.

[\textbf{MM}] \ K. B. Marathe and G. Martucci, \textit{The Mathematical
Foundations of Gauge Theories}, North-Holland, Amsterdam, 1992.

[\textbf{M}] \ C. B. Morrey, Jr., \textit{Multiple Integrals in the Calculus
of Variations}, Springer-Verlag, Berlin, 1966.

\textbf{[O1]} \ T. H. Otway, Yang-Mills fields with nonquadratic energy, 
\textit{J. Geometry \& Physics} \textbf{19 }(1996),379-398.

\textbf{[O2]} \ T. H. Otway, Properties of nonlinear Hodge fields, \textit{%
J. Geometry \& Physics }\textbf{27 }(1998), 65-78.

[\textbf{O3}] \ T. H. Otway, Nonlinear Hodge maps, \textit{J. Math. Physics} 
\textbf{41} (2000), 5745-5766.

[\textbf{O4}] \ T. H. Otway, Nonlinear Hodge structures in vector bundles,
in: \textit{Nonlinear Analysis in Geometry and Topology,} Th. M. Rassias,
ed., Hadronic Press, Palm Harbor, 2000; see also arXiv:math-ph/9808013.

[\textbf{O5}] \ T. H. Otway, Nonlinear Hodge maps and rotational fields,
preprint.

[\textbf{P}] \ P. Price, A monotonicity formula for Yang-Mills fields, 
\textit{Manuscripta Math.} \textbf{43} (1983), 417--491.

[\textbf{Se}] \ J. Serrin, Removable singularities of solutions of elliptic
equations, \textit{Archs. Rat. Mech. Anal.} \textbf{17} (1964), 67-78.

\textbf{[Si]} \ L. M. Sibner, An existence theorem for a nonregular
variational problem, \textit{Manuscripta} \textit{Math.} \textbf{43 }(1983),
45-72.

\textbf{[SS1]} \ L. M. Sibner and R. J. Sibner, A nonlinear Hodge-de Rham
theorem, \textit{Acta Math.} \textbf{125 }(1970), 57-73.

\textbf{[SS2]} \ L. M. Sibner and R. J. Sibner, Nonlinear Hodge theory:
Applications, \textit{Advances in Math.} \textbf{31} (1979), 1-15.

[\textbf{T}] \ S. Takakuwa, On removable singularities of stationary
harmonic maps, \textit{J. Fac. Sci. Univ. Tokyo, Sect. 1}, \textbf{32}
(1985), 373-395.

\textbf{[U1]} \ K. Uhlenbeck, Regularity for a class of nonlinear elliptic
systems, \textit{Acta Math.} \textbf{138 }(1977), 219-240.

\textbf{[U2] \ }K. Uhlenbeck, Removable singularities in Yang-Mills fields, 
\textit{Commun. Math. Physics} \textbf{83 }(1982), 11-30.

[\textbf{W}] \ \ H. Wallin, A connection between $\alpha $-capacity and $%
L^{p}$-classes of differentiable functions, \textit{Ark. Mat.} \textbf{5},
No. 24 (1964), 331-334.

[\textbf{Y}] \ Y. Yang, Classical solutions in the Born-Infeld theory, Proc.
R. Soc. Lond. Ser. A \textbf{456} (2000), no. 1995, 615-640.

\end{document}